\documentclass[aip,apl,twocolumn,reprint]{revtex4-1}
\usepackage{graphicx}
\usepackage{dcolumn}
\usepackage{bm}
\usepackage{color}
\usepackage{tabularx}
\usepackage{array}
\usepackage[utf8]{inputenc}
\usepackage{color,soul}
\usepackage{amsmath}

\draft 

\begin{document}


\title{Modulating Surface Acoustic Wave Generation through Superconductivity} 



\author{Andrew Christy}
\thanks{achri@unc.edu}
\affiliation{Department of Physics and Astronomy, University of North Carolina at Chapel Hill, Chapel Hill, NC 27599, USA}
\affiliation{Department of Chemistry, University of North Carolina at Chapel Hill, Chapel Hill, NC 27599, USA}

\author{Yuzan Xiong}
\affiliation{Department of Physics and Astronomy, University of North Carolina at Chapel Hill, Chapel Hill, NC 27599, USA}

\author{Rui Sun}
\affiliation{Department of Physics and Organic and Carbon Electronics Lab (ORaCEL), North Carolina State University, Raleigh, NC 27695, USA}

\author{Yi Li}
\affiliation{Materials Science Division, Argonne National Laboratory, Argonne, IL 60439, USA}

\author{Kenneth O. Chua}
\affiliation{Department of Chemistry, University of North Carolina at Chapel Hill, Chapel Hill, NC 27599, USA}

\author{Andrew H. Comstock}
\affiliation{Department of Physics and Organic and Carbon Electronics Lab (ORaCEL), North Carolina State University, Raleigh, NC 27695, USA}

\author{Junming Wu}
\affiliation{Department of Physics and Astronomy, University of North Carolina at Chapel Hill, Chapel Hill, NC 27599, USA}

\author{Sidong Lei}
\affiliation{NanoScience Technology Center, University of Central Florida, Orlando, FL 32826, USA}
\affiliation{Department of Material Science and Engineering, University of Central Florida, Orlando, FL 32816, USA}

\author{Frank Tsui}
\affiliation{Department of Physics and Astronomy, University of North Carolina at Chapel Hill, Chapel Hill, NC 27599, USA}

\author{Megan N. Jackson}
\affiliation{Department of Chemistry, University of North Carolina at Chapel Hill, Chapel Hill, NC 27599, USA}

\author{Dali Sun}
\affiliation{Department of Physics and Organic and Carbon Electronics Lab (ORaCEL), North Carolina State University, Raleigh, NC 27695, USA}

\author{Valentine Novosad}
\affiliation{Materials Science Division, Argonne National Laboratory, Argonne, IL 60439, USA}

\author{James F. Cahoon}
\affiliation{Department of Chemistry, University of North Carolina at Chapel Hill, Chapel Hill, NC 27599, USA}

\author{Wei Zhang}
\thanks{Correspondence should be addressed to: zhwei@unc.edu}
\affiliation{Department of Physics and Astronomy, University of North Carolina at Chapel Hill, Chapel Hill, NC 27599, USA}

\date{\today}

\begin{abstract}
Surface acoustic waves (SAWs), with their five orders-of-magnitude slower propagation velocity, allow for considerably shorter wavelengths at the same frequency compared to electromagnetic waves. The short wavelengths allow for device miniaturization and on-chip integration. The generic design of these devices involve piezoelectric substrates with comblike arrays of Al or Au electrodes known as interdigitated transducers deposited on the surface. \textcolor{black}{However, Al and Au both have shortcomings at the cryogenic temperatures required for quantum applications, namely the formation of two-level systems and the lack of superconductivity perpetuating Ohmic losses, respectively.} In this work, SAWs are generated in the high-MHz to low-GHz range using niobium nitride (NbN) interdigitated transducers (IDTs) and Bragg reflectors. We demonstrate the fabrication of acoustic devices through photolithography and reactive ion etching (RIE). \textcolor{black}{The sharp transition between superconducting and normal states and the corresponding change in SAW transmission allows for fine control of the "on" (superconducting) and "off" (normal) states of NbN, with a $\Delta T=$1 K separating the transmission minimum and maximum. We demonstrate a 16$\times$ difference in transmission between the "on" and "off" states of the device.} The SAW transmission behavior mirrors the change in resistance of NbN at its $T_c$. \textcolor{black}{These findings open up new possibilities for the integration of NbN SAW resonators into existing quantum architectures based on NbN and a method for adjusting transmission properties independent of applied voltage.}
\end{abstract}

\pacs{}

\maketitle 


Hybrid phonon polarons have recently attracted renewed interests due to their applications in hybrid quantum technology \cite{satzinger2018quantumcontrol, manenti2017cqad, magnusson2015zno,awschalom2025challenges}, in which the phonon excitations coherently couple to other excitations such as magnons in ferromagnets \cite{hwang2024stronglycoupled, xu2020nonreciprocity, shah2020nonreciprocity}, excitonic in 2D materials \cite{huang2017mos2, preciado2015mos2, rezk2016mos2}, and single spin in solid defects \cite{lemonde2018vacancy, golter2016nivacancy, golter2016coupling, whiteley2019gaussian}. The unique properties of phonons such as their slow propagation velocity and long lifetimes in solid-state platforms \cite{awschalom2021quantum} render them a promising candidate for building hybrid quantum systems. Surface acoustic waves (SAWs) are especially appealing for applications such as these, due to their confinement at the surface of the material for interactions with thin films deposited on the substrate \cite{weiler2011fmr} and convenient wavelength and frequency tunability from engineering the IDTs. The surface confinement also allows the wave to be accessed along its propagation path, and greatly broadens the applications of acoustic devices as chemical, physical, and biological sensors and transducers. The accurate control of the phonon wavelength and high Q-factor also render them widely applicable for rf and mircrowave circuit components, including but not limited to, delay lines, oscillators, resonators, and bandpass filters \cite{weiyi2023integrated,yu2022integrated,zhao2022enabling}.  

A notable development in this field is circuit quantum acoustodynamics (cQAD), where a superconducting qubit is coupled to high-MHz or low-GHz acoustic modes \cite{manenti2017cqad}, i.e., a solid-state analog of cavity quantum electrodynamics (cQED), where the interaction between a qubit and microwave resonator allows for control and readout of quantum states on a chip \cite{wallraff2004cqed}. The cQAD permits the piezoelectric coupling of a qubit to an acoustic cavity, which can be customized via appropriate designs of interdigital transducers (IDTs) and/or Bragg reflectors. To date, both SAW and bulk acoustic wave (BAW) resonators have been used to show strong interactions with superconducting qubits \cite{von2024engineering}. The SAW cavities at these frequencies still show good performance when cooled to cryogenic temperatures, where suspended mechanical resonators suffer high losses \cite{magnusson2015zno}.

\begin{figure*}[htb]
 \centering
 \includegraphics[width=6.2 in]{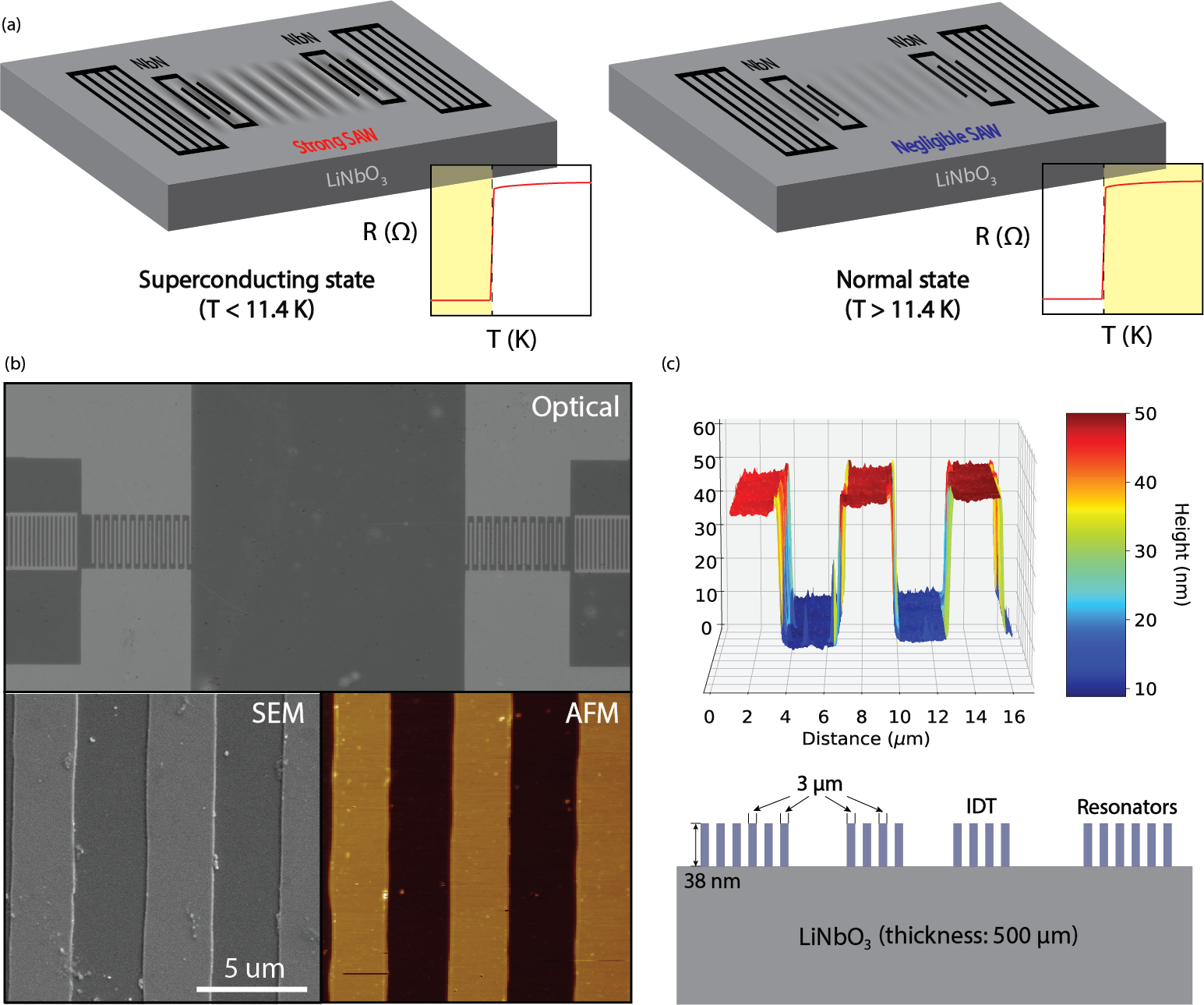}
 \caption{(a) Schematic illustration of the NbN SAW resonator on lithium niobate. Above the $T_c$ of NbN, SAW transmission is negligible. Below the T$_c$, the superconducting nature of NbN induces strong SAW transmission. (b) Optical, SEM, and AFM images of the SAW device. \textcolor{black}{The electrode edge roughness is due to limited resolution in photolithography}. (c) Surface morphology profile of the SAW device scanned by AFM along with a side view of the device design including the thicknesses of the LNO (500 $\mu$m) and NbN (38 nm).}
 \label{fig1}
\end{figure*}

Acoustic devices such as these have been built on piezoelectric crystal or film materials such as quartz \cite{manenti2017cqad}, aluminum nitride \cite{chu2017aln}, zinc oxide \cite{golter2016nivacancy}, and lithium niobate \cite{xu2020nonreciprocity, hwang2020cavity}. SAWs generated on these materials travel at a velocity of $\approx$ 3000 m/s \cite{datta1986SAW}, five orders of magnitude slower than the speed of light. Electromagnetic microwave devices in the highly technologically relevant GHz regime are generally bulky, since the corresponding wavelength of light ranges between cm and m \cite{delsing20192019}. This vast difference in velocity allows for miniaturization of SAW devices past the lower size limit of photonic circuits. SAW devices have demonstrated advantages in quantum signal processing, as the slow propagation speed allows multiple wavelengths (signals) to be manipulated on a single chip \cite{manenti2017cqad}. 
The traveling SAW modes can be excited by applying an oscillating voltage to the IDT electrodes. The Q-factors of these devices have been shown to be comparable to the cQED device counterparts \cite{shao2020integrated, shao2022electrical}. 

Whereas previous research has used metallic IDTs almost exclusively from Al or Au \cite{weiler2011fmr, hwang2020cavity}, here, we present IDTs fabricated with superconducting niobium nitride (NbN). NbN is a well-understood superconducting material, with an easily accessible critical temperature ($\approx$ 11 K for this geometry), high critical field, good mechanical strength, stability in atmospheric conditions, and no observable hysteresis when moving between cryogenic and room temperatures \cite{chockalingam2008nbn}. Studies of its superconductivity have been motivated by applications in Josephson junctions \cite{kawakami2003development}, low noise hot electron bolometer mixers \cite{hajenius2004nbn}, and single-photon detectors \cite{korneev2005nbn}. \textcolor{black}{At the milikevin (mK) temperatures required for quantum applications, both Au and Al have well documented shortcomings. Au does not have an experimentally realized superconducting state \cite{ummarino2024can}, therefore maintaining Ohmic losses even at cryogenic temperatures. For Al, the Al$_2$O$_3$ oxidative barrier that forms on the surface of Al transducers can introduce two-level systems (TLS) that contribute to qubit decoherence \cite{gordon2014hydrogen}. The use of NbN/AlN/NbN trilayers has already been demonstrated to bypass the TLS problem in superconducting quantum circuits \cite{kim2021enhanced}, but this new design paradigm has yet to be extended to quantum acoustics.}

Here, we demonstrate the use of NbN IDTs in a 2-port SAW resonator. We characterize the behavior of the cavity through rigorous time- and frequency-domain analysis, and demonstrate control of SAW properties through the use of superconductivity. \textcolor{black}{We show that NbN IDTs lack the internal reflections that cause a spurious triple transit signal and significant warping to the frequency response. Because of this, our device can be accurately modeled by a simple quasi-static delta function model, without the extra complications that come with electrode reflectivity. Currently, the triple transit signal is minimized by either accepting a large insertion loss \cite{morgan2007saw} or by altering the device geometry in ways that lower the operating frequency \cite{ekstrom2017surface, weiss2018multiharmonic}}. By using NbN IDTs, it is possible to implement a SAW platform that can be easily integrated into existing nitride-based quantum architectures \cite{kim2021enhanced}. At the temperatures required for the function of superconducting qubits and circuits, the transmission behavior of NbN IDTs can be conveniently optimized. 

\begin{figure*}[htb]
 \centering
 \includegraphics[width=6.3 in]{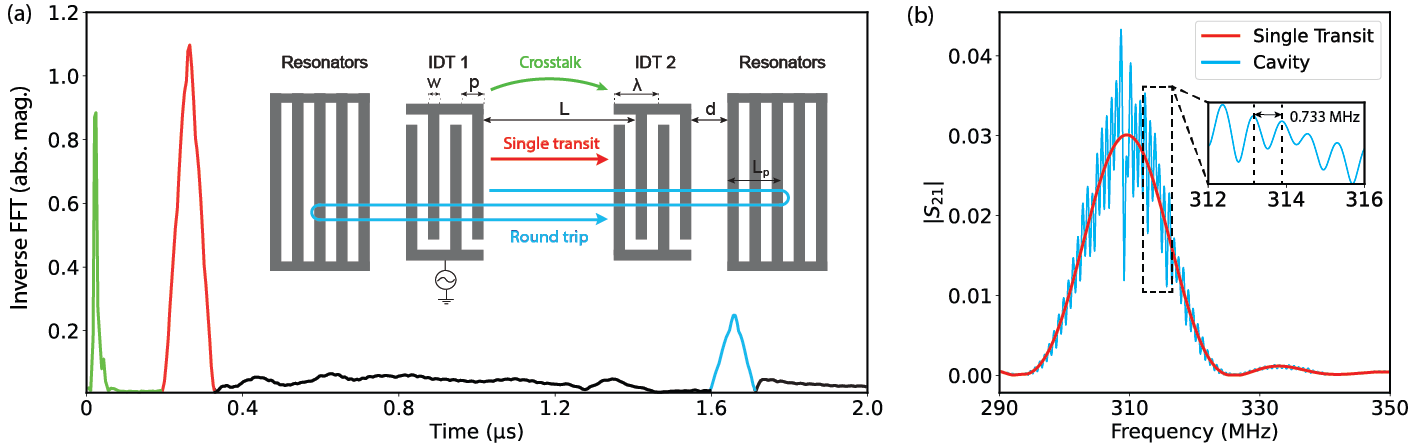}
 \caption{(a) Time domain results of the NbN IDT device measured at 11 K. The free space electromagnetic wave(EMW) occurs at 10.7 ns, the single-transit SAW pulse (red) occurs at 178.6 ns, and the round-trip SAW pulse (blue) occurs at 1.541 $\mu$s. The inset shows the device geometry, with dimensions \textit{L} = 618 $\mu$m, \textit{d$_{T}$} = 225 $\mu$m, \textit{d} = 36 $\mu$m. The IDT dimensions are \textit{w} = 3 $\mu$m, \textit{p} = 6 $\mu$m, \textit{$\lambda$} = 12 $\mu$m, designed for maximum fundamental frequency amplitude. (b) Frequency domain S$_{21}$ results of the device measured at 11 K for the single-transit SAW pulse (red) and the full cavity SAWs (blue). The fine splitting is due to the cavity that modulates the overall sinc$^2$ profile of the SAW mode. \textcolor{black}{The inset shows the determination of the FSR value of the cavity, 0.733 MHz.}}
 \label{fig2}
\end{figure*}

For the present study, we used $128^{\circ}$ $Y$-$X$ cut lithium niobate (LNO) as our piezoelectric substrate, due to its strong electromechanical coupling and low loss. Figure \ref{fig1}(a) schematically illustrates the geometry of our device, which includes both IDTs and Bragg mirrors made from superconducting NbN. \textcolor{black}{Below the measured $T_c$ of our device, while the NbN is in its superconducting state, the IDTs generate and receive strong SAWs. Above the $T_c$, the SAW transmission is negligible. In the fabrication process, the planar NbN film was first deposited using ion-beam-assisted, room-temperature reactive sputtering \cite{Polakovic18}, which yields a high $T_c$ of 12.6 K. The difference in pristine film $T_c$ and device $T_c$ will be discussed later.} The surface acoustic wave structures were fabricated using photolithography and reactive ion etching, which have been shown to provide minimal edge damage to the microwave transmissions \cite{LiPRL19,LiPRL22}. Figure \ref{fig1}(b) shows the IDT and Bragg reflector array inspected under optical microscopy, scanning electron microscopy (SEM), and atomic force microscopy (AFM). \textcolor{black}{The actual thickness of the patterned NbN film is determined as 38 nm by the AFM. } The IDTs consist of 19 finger pairs, which have a voltage applied to them through NbN contact pads. \textcolor{black}{The fingers have a designed width $w$ of 3 $\mu$m, and a pitch $p$, defined as the distance between the centers of adjacent fingers, of 6 $\mu$m, for $w/p = 0.5$.} This ``metallization ratio'' also known as $\eta$, determines the relative intensity of the fundamental frequency and higher harmonic peaks, and a $\eta$ of 0.5 is used to maximize the intensity of the fundamental mode \cite{datta1986SAW}. Often, due to the slight nonuniformity in the lithography process, slight deviations from the theoretically-calculated SAW wavelength, $\lambda_\textrm{SAW}$ (and hence the frequency response) of the devices may occur. However, due to the large number of finger pairs and reflectors used in the devices, irregularities as such are mostly averaged out. The quantitative effect of the frequency response deviations will be discussed later. To resonantly enhance the SAW transmission, a set of 152 NbN Bragg reflectors were fabricated on either side of the IDTs. The reflectors are strips of NbN that form a shorted IDT, with a spacing of $\lambda_\textrm{SAW}/2$. Since the single-finger IDT design is bidirectional, these mirrors serve to reflect the SAWs that are generated traveling the opposite direction from the receiver IDT, as well as any SAWs that leak through the receiver IDT. Since the reflectors have the same $w$ and $p$ as the IDT electrodes, the cavity is tuned to reflect the SAW signals at the same wavelength being generated by the IDT. Additionally, since the distance $d$ between the IDT and reflector array is an integer multiple of $\lambda_\textrm{SAW}$, the reflected SAWs constructively interfere and lead to a net enhancement of the transmission. Figure \ref{fig1}(c) shows atomic force microscopy (AFM) measurements of the device surface. The average height of the IDTs and Bragg reflectors was 38 nm, which is slightly less than the thin film from which this device was etched. 

The measurement was performed in the frequency domain with a commercial vector network analyzer (VNA, Anritsu MS46322B). The time domain S$_{21}$ was calculated using an inverse fast Fourier Transform (IFFT) of the frequency domain S$_{21}$ from 200 - 420 MHz, with a Hanning window function and zero padding procedure. Figure \ref{fig2}(a) shows the time domain data for the device measured at 11 K, along with a schematic of the device and pertinent SAW modes in the inset. The red peak corresponds to the single-transit SAW mode, where the propagation distance is just the gap between the generating and receiving IDTs. The width of the peak is proportional to the dimensions of the IDT and the number of finger pairs. The blue peak corresponds to the round-trip SAW mode, whose propagation distance is the gap between IDTs plus two times the cavity length \cite{ekstrom2017surface}. The time delays of these peaks can be used to extract parameters of the device such as surface velocity, cavity length, and penetration distance of the Bragg reflector array. The frequency domain result at 11 K is shown in Fig. \ref{fig2}(b). The single-transit SAW mode has a $\textrm{sinc}^2$ profile, which then undergoes fine splittings due to the cavity. \textcolor{black}{Fitting the red curve in Fig. \ref{fig2}(b) to a $\textrm{sinc}^2$ function gives a resonance frequency of 309 MHz.} The relationship between Fig. \ref{fig2}(a) and Fig. \ref{fig2}(b) is a Fourier transform, where the single-transit SAW mode is a transform of the red peak, and the cavity trace is a transform of everything after the onset of the red peak. Therefore, the cavity trace encapsulates all of the SAW reflections due to the Bragg reflector array.

\begin{figure*}[htb]
 \centering
 \includegraphics[width=6.3 in]{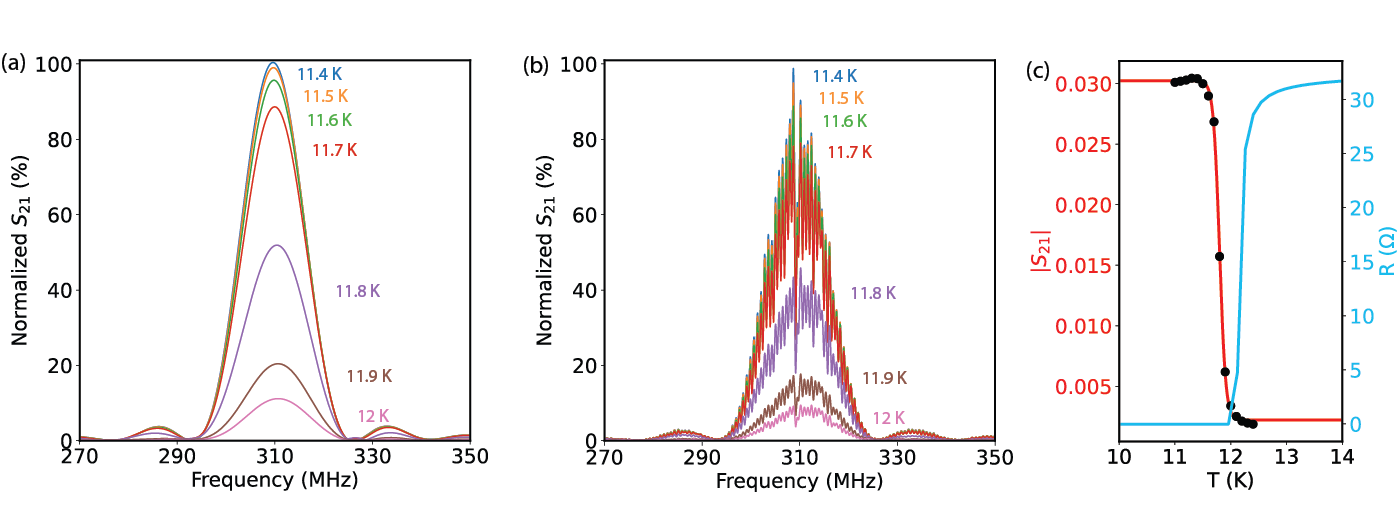}
 \caption{(a) S$_{21}$ spectrum of the single-transit SAW at the fundamental frequency over the superconducting transition (11.4 - 12 K). (b) S$_{21}$ of the cavity SAW at the fundamental frequency over the same temperature range. (c) The temperature dependence of the S$_{21}$ at the fundamental frequency (black dots), fitted to a logistic function (red). The blue trace is a separate resistance measurement of a 40-nm continuous NbN film.}
 \label{fig3}
\end{figure*}

The temperature dependence of both the single-transit SAW and the cavity SAW are shown in Fig. \ref{fig3}(a) and (b). The maximum transmission occurs at 11.4 K at 309 MHz, and levels off when the temperature is decreased below 11.4 K. The single-transit SAW and cavity SAW have the same sinc$^2$ profile and the fine splitting feature as shown in Fig. \ref{fig2}(b). Figure \ref{fig3}(c) shows the change in maximum $S_{21}$ at 309 MHz over the assumed temperature range for the superconducting transition of the NbN. Above $T_c$, the device measures a 30-$\Omega$ resistance that hinders its ability to conduct the necessary current to piezoelectrically excite SAWs in the lithium niobate. However, as the temperature decreases to below $T_c$, the resistance drops to zero (superconducting state), which consequently results in a sudden, large increase in the transmission. The resistance measurement for a reference film sample of 40-nm NbN grown on silicon is also shown in Fig. \ref{fig3}(c). The temperature dependence of the transmission character matches quite well with that of the resistance measurement. The slight difference in the $T_c$ measured by either method (resistance or transmission) is attributed to the difference in the geometries of the samples (planar film versus device): the edge roughness due to the etching process during fabrication and the comblike array geometry could cause the $T_c$ of the device to decrease as compared with the continuous, uniform film \cite{zgirski2005size}. \textcolor{black}{The fit to the transmission at 309 MHz over the temperature range, as shown by the red trace in Fig.\ref{fig3}(c), levels off at 6$\%$ of the maximum transmission at 11.4 K. This shows the 16$\times$ transmission difference between the "on" and "off" states of the device.} While the piezoelectric properties of materials have a temperature dependence, LNO is well regarded to maintain its stability at cryogenic temperatures \cite{lin2024simple}, so we attribute the transmission changes to the behavior of the NbN. \textcolor{black}{The 5\textsuperscript{th} harmonic is also visible at 1.55 GHz and has an identical temperature dependence to the fundamental frequency.}     

Apart from the frequency domain analysis, the time delays of the SAW pulses are necessary to characterize the behavior of both the IDT and the Bragg reflectors. We analyzed the time domain behavior of the cavity using the device dimensions and physical parameters of the substrate, and applied a mathematical model of the frequency response of a 2-port SAW device. We further simulated the signal transmission parameters of an VNA using these parameters for a direct comparison to the experimental results discussed above.
 
Since the SAWs must first be excited by an electrical signal applied to the IDT fingers, the launch time for the SAW is equal to the time that the EMW reaches the receiver IDT (the EMW travel time between the IDTs since the distance is negligible compared to the traveled distance through the cables). Therefore, the travel time \textit{t} is equal to $t_\textrm{SAW} - t_\textrm{EMW}$, with $t_\textrm{EMW} = 10.7$ ns for this measurement setup. The free substrate velocity for LNO can be determined by using the travel time of the leading edge of the single-transit SAW shown in Fig.\ref{fig2}(a). The SAW leading edge only travels the distance $L$ between the IDTs, consequently bypassing any small velocity-decreasing effects of the electrode loading, $v_p = \frac{L}{t_\textrm{SAW} - t_\textrm{EMW}}$, with $t_\textrm{SAW} = 178.6$ ns. This gives a $v_p$ of 3716 m s$^{-1}$, which is less than the literature value for $v_p$ along the crystal $X$ axis at room temperature of 3950 m s$^{-1}$  \cite{datta1986SAW}. We attribute the difference to either a slight deviation off the $X$ axis, which would decrease the speed of the SAW, or crystal contractions at cryogenic temperatures \cite{manenti2017cqad}. We also note the absence of a triple transit signal, which would be a pulse of the same width as the single transit SAW at 547 ns. This eliminates extra ripples that would show up in the frequency response and tells us that the splitting is wholly due to the reflector cavity. \textcolor{black}{We can conclude that the triple transit signal is at least reduced below the noise floor, but a determination of whether the signal is just reduced to a trivial amount or totally suppressed due to superconductivity merits future systematic study.}   

Each individual Bragg reflector reflects a small portion of the SAW, with an additive effect over the whole array. However, for simplification, the array can be represented as a mirror set at a penetration distance, up to which the leftward-traveling SAW travels and is collectively reflected back towards the receiver IDT \cite{morgan2007saw}. The penetration distance $L_p$ and the reflection coefficient $|r_s|$ can be found using the time delay of the second large pulse in the cavity device result, occurring at 1.541 $\mu$s. This effect is apparent in Fig.\ref{fig2}(a), where the pulse loses amplitude due to losses from multiple reflections. This second pulse is the round-trip SAW pulse, which has traveled a distance of $2L_c + L$.  

The time delay between the single-transit SAW and round-trip SAW is 1.363 $\mu$s, which corresponds to a path length difference of $\Delta L$ = 5064 $\mu$m for SAWs traveling at $v_p$. Based on the geometry of the cavity, $\Delta L = 2L_c$. Therefore, $L_c$ = 2532 $\mu$m and $L_p$ = 696 $\mu$m. The single strip reflection coefficient $|r_s|$ = $\frac{w}{L_p}$ = 0.0043. Measurements of analogous Al cavities on LNO show a single strip reflection coefficient of 0.03 - 0.04 (see Supplemental Materials \cite{supplementalmaterial}), which are similar to literature values \cite{manenti2017cqad}. \textcolor{black}{Reflective transducers are partly the cause of the triple transit signal, so a lower single strip reflection coefficient leads to a less intense triple transit.} \textcolor{black}{By fitting the decay of pulse amplitude over time to an exponential $A = A_0e^{\frac{-\pi f_0 t}{Q}} $, we are able to calculate a Q-factor of 2220 for our acoustic cavity \cite{nawrodt2008high}}. 

\textcolor{black}{In general, SAW cavities benefit from high electrode reflectivity, due to the confinement of acoustic energy for higher Q-factors \cite{gokhale2020epitaxial}. However, the low reflection coefficient of NbN eliminates the internal reflections in the IDT itself, leading to a symmetric frequency response \cite{morgan2007saw}. Currently, split finger transducer designs are used to eliminate the reflectivity inherent in all metal IDTs \cite{weiss2018multiharmonic}. This structure halves the operating frequency of a single finger IDT at the same electrode width and pitch, but the use of NbN transducers eliminates the reflections in the single finger architecture, allowing the single finger design to obtain the benefits of the split finger. While our device does not reach the high Q-factors of state-of-the-art acoustic cavity designs \cite{shao2019phononic}, we believe that the low reflectivity of the IDTs causes fewer resistive losses that would hamper the Q-factor even more. } 

Our acoustic cavity supports a large number of modes, each separated by the free spectral range (FSR) \cite{hecht2002optics} of the cavity and can normally be estimated by $\frac{v_p}{2L_c}$, where $L_c = d_{total} + 2L_p$ with $d_{total}$ being the distance between the gratings \cite{hwang2024stronglycoupled}. For this device, $L_c$ = 2532 $\mu$m, corresponding to a FSR of 0.734 MHz. The average spacing between each adjacent peak in Fig.\ref{fig2}(b) is 0.733 MHz, which is consistent with the cavity length and FSR found using time domain analysis.

\begin{figure*}[htb]
 \centering
 \includegraphics[width=6.3 in]{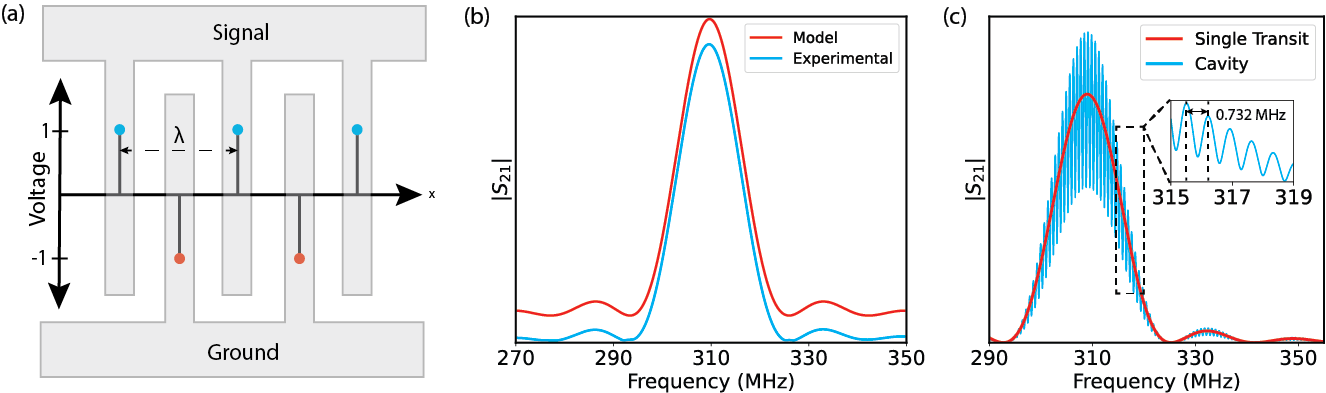}
 \caption{(a) The line charge model representation of the IDTs in a SAW cavity. (b) Comparison of the normalized experimental data and calculated results from Eq.\ref{eq01} with a vertical offset for visual clarity. (c) Simulated frequency domain data, with both single-transit and cavity SAW modes contributions. \textcolor{black}{The inset shows the determination of the FSR value of the cavity, 0.732 MHz.}}
 \label{fig4}
\end{figure*}

The IDT can be mathematically modeled by discretizing the fingers as line charges. A visual representation of such is shown in Fig.\ref{fig4}(a). Each finger has a finite width, but is simplified to a line charge at its center. The distance between fingers of similar polarity is $\lambda$, the wavelength of the IDT. The distance between fingers of opposite polarity is $p$, the pitch of the IDT ($\lambda /2$). Thus the frequency response is written as:  

\begin{equation}
X_f = \sum_{n = 0}^{N_e - 1} (-1)^n e^{-i\pi n f \frac{\lambda}{v_p}}
\label{eq01}
\end{equation}
Since this is the frequency response of a single IDT, and the device consists of both an exciter and reciever IDT (assumed to be identical), the transmission S$_{21}$ is equal to $|X_f|^2$. This closely matches a sinc$^2$ function \cite{campbell1989saw}, and it can be understood by the fact that the frequency response of an IDT is the Fourier transform of its impulse response, which is represented by a rectangular function and discretized by the $\delta$-function line charges. The resulting frequency response then takes the form: 

\begin{equation}
\label{eq02}
|X_{IDT}(f)|^2 = \Bigg ( N_e \Bigg |\frac{\sin{(N\pi (f - f_0)/f_0)}}{N\pi (f - f_0)/f_0} \Bigg | \Bigg ) ^2
\end{equation}
\noindent where $f_0 = v_p /\lambda$ is the center frequency of the IDT, $N$ is the number of finger pairs, and $N_e$ is the total number of electrodes ($2N$).

Fig. \ref{fig4}(b) shows the comparison between the modeled data calculated by \ref{eq01} using v$_p$ = 3716 m/s and $\lambda = 12$ $\mu$m and the experimental single transit SAW pulse. The center frequencies differ by only 109 kHz, which confirms the fidelity of both our calculations and our fabrication method. 

We developed a simulation of the transmission parameter S$_\textrm{21}$ using the line-charge model, which allows a direct comparison with the experimental data acquired using the VNA. The sinc$^2$ output of the model was added to an electromagnetic background corresponding to the VNA. The details of the simulation can be found in \cite{supplementalmaterial}. For simplicity, the reflector array is represented as a mirror at $L_p$. The time delays of the single-transit and round-trip SAWs are calculated using the device parameters and calculated surface wave velocity. Using the dimensions of the device in Fig. \ref{fig2}(a) and the calculated single strip reflection coefficient $|r_s|$, the model predicts a frequency spacing of 0.732 MHz. The output of the simulation, in frequency domain, can be seen in Fig. \ref{fig4}(c). This FSR value matches well with both the experimental and the simulated values.

In summary, we investigated the use of a superconducting material to fabricate surface acoustic wave devices, demonstrating a temperature-based modulation of wave propagation. We showed that below the $T_c$ of NbN, our IDTs produce strong SAW signals, whereas such SAW generation is suppressed when the material is brought to a non-superconducting state. We demonstrated a 16$\times$ increase in transmission as NbN goes from its normal state to the superconducting state. We thoroughly characterized the behavior of the fabricated acoustic cavity, and demonstrated a NbN single-strip reflection coefficient that is an order of magnitude lower than the standard Al. 

\textcolor{black}{Our experimental results and theoretical modeling indicate that the NbN in its superconducting state performs better as an IDT material due to its lack of resistive losses and reflections, but poorly as a Bragg reflector material for this same reason. We believe that a device design that integrates NbN IDTs and Au/Al reflectors could perform better than an all-NbN or all-metal device. We also believe that more modeling is necessary to understand the almost perfectly symmetrical frequency response of NbN IDTs. In the future, we hope to expand on work using all-nitride superconducting qubits \cite{kim2021enhanced} and integrate them with our niobium nitride SAW resonator}. Overall, these findings represent a step forward in using superconducting acoustic cavities for quantum applications. Our characterization of NbN and its behavior in an acoustic cavity also pave the way for its integration into current cQAD architectures to take advantage of its unique tunabilities. 

\section*{Supplementary Material}

See for cavity description and characterization of additional Al devices, details of the VNA output and data modification simulation, and description of the anomalous generation of BAWs.

\begin{acknowledgments}
The experimental work at UNC-CH was primarily supported by the U.S. Department of Energy (DOE), Office of Science, Basic Energy Sciences under award number DE-SC0026305. Part of the work related to the data curation and modeling was supported by the NSF under award No. DMR-2509513. This work was performed in part at the Chapel Hill Analytical and Nanofabrication Laboratory, CHANL, a member of the North Carolina Research Triangle Nanotechnology Network, RTNN, which is supported by the National Science Foundation, Grant ECCS-2025064, as part of the National Nanotechnology Coordinated Infrastructure, NNCI. S.L. acknowledges support from the National Science Foundation (NSF) under Grant No. DMR-2209427. D.S. acknowledges support from the NSF under Grant No. DMR-2143642. Y.L. is supported by the U.S. DOE, Office of Science, Basic Energy Sciences, Materials Science and Engineering Division under contract No. DE-SC0022060.
\end{acknowledgments}


\bibliography{references}

\end{document}